\documentclass[runningheads]{llncs}
\usepackage{graphicx}
\usepackage{hyperref}
\usepackage{wrapfig}

\begin{document}

\title{``GAN I hire you?'' - A System for Personalized Virtual Job Interview Training}


%
\titlerunning{Personalized Behavioral Feedback through Adversarial Learning}
%
\author{Alexander Heimerl\inst{1} \and
Silvan Mertes\inst{1} \and
Tanja Schneeberger\inst{2} \and Tobias Baur\inst{1} \and Ailin Liu\inst{1} \and Linda Becker\inst{3} \and Nicolas Rohleder\inst{3} \and Patrick Gebhard \inst{2} \and Elisabeth Andr\'{e}\inst{1}}

\institute{Lab for Human-Centered AI, Augsburg University, Augsburg 86159, Germany
\email{\{alexander.heimerl, silvan.mertes, tobias.baur, elisabeth.andre\}@uni-a.de}
\and German Research Center for Artificial Intelligence (DFKI), 
Saarland Informatics Campus D3.2, Saarbrücken, Germany
\email{\{schneeberger, gebhard\}@dfki.de}
\and Department of Psychology, Friedrich-Alexander University Erlangen-Nürnberg, Erlangen 91054, Germany
\email{\{linda.becker, nicolas.rohleder\}@fau.de}}

\authorrunning{Heimerl and Mertes et al.}
%
\maketitle              
\begin{abstract}
Job interviews are usually high-stakes social situations where professional and behavioral skills are required for a satisfactory outcome.
Professional job interview trainers give educative feedback about the shown behavior according to common standards. This feedback can be helpful concerning the improvement of behavioral skills needed for job interviews. A technological approach for generating such feedback might be a playful and low-key starting point for job interview training. 
Therefore, we extended an interactive virtual job interview training system with a Generative Adversarial Network (GAN)-based approach that first detects behavioral weaknesses and subsequently generates personalized feedback. 
To evaluate the usefulness of the generated feedback, we conducted a mixed-methods pilot study using mock-ups from the job interview training system.
The overall study results indicate that the GAN-based generated behavioral feedback is helpful. Moreover, participants assessed that the feedback would improve their job interview performance.










\end{abstract}
\section{Introduction}
\label{sec:introduction}
Given the global economic situation, we are confronted with a (post-) Covid-19 pandemic world, one significant issue many countries around the globe face is the rising number of people Not in Employment, Education, or Training (NEETs). NEETs often suffer from underdeveloped socio-emotional and interaction skills \cite{macdonald2008disconnected,hammer2000mental}, such as a lack of self-confidence and perception of one's strengths. This circumstance might affect their performance in various essential situations, such as job interviews. They need to convince a recruiter of their fit in a company by actively engaging in interviews. Interviewers consciously or unconsciously heavily rely on social cues to assess the fitting. The amount of positive engagement a candidate shows towards the interviewer may play a central role in whether the candidate is suitable. Delroy et al.\cite{delroy:2013:jobinterview_role_personality_culture} found that active integration behaviors such as engagement, laughing, and humor led to better performance ratings and, therefore, to a higher rating of overall recruitability. As many people lack such social skills, many countries have specialized inclusion centers before the pandemic, meant to aid people to secure employment through coaching by professional practitioners. However, this approach is costly and time-consuming, and given lockdown situations often cannot be performed physically. In this context, technology-based solutions might be reasonable alternatives to the existing human-to-human coaching practices.

This paper presents a feedback extension to an existing job interview training environment that uses a socially interactive agent as a recruiter and an engagement recognition component to enable the virtual agent to react and adapt to the user's behavior, and emotions \cite{baur2015}. 
There, during a preparation phase, trainees were instructed to show certain behaviors in specific job interview training situations and got feedback from the virtual agent on whether they could perform these instructions correctly. 
This training aims to help improve social skills that are pertinent to job interviews. The new feedback extension employs an eXplainable AI (XAI) method based on counterfactual reasoning for generating verbal feedback about observed social behavior. This approach allows communicating features (e.g., no eye contact, closed body posture) that weaken the overall job interview performance. There, we make use of \emph{counterfactual explanations}, explaining to a user that a modified version of her/his social behavior would have led to a better behavior rating. In the context of XAI, counterfactual explanations have proven their capabilities to lead to a reasonable explanation satisfaction, since the user is not only presented with information about \emph{which} features are relevant but further, even if implicitly, with information about \emph{why} those features are relevant. 

\begin{figure}[ht]
\vspace{-15pt}
\centering
\includegraphics[width=1\textwidth]{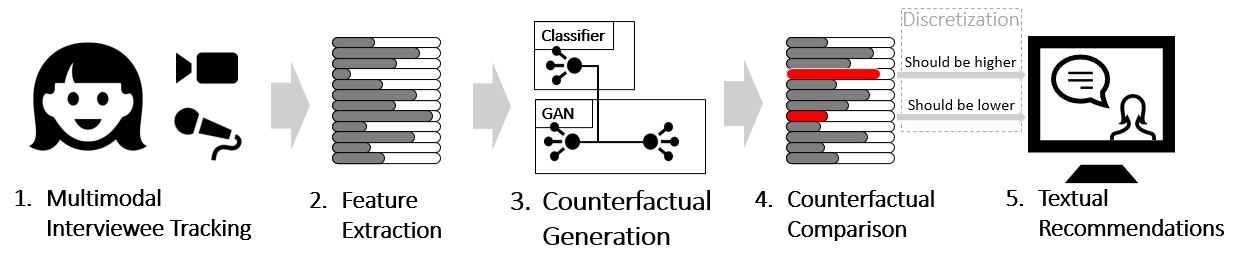}
\caption{Job interview training system with GAN-generated recommendations.}
\label{fig:schema}
\end{figure}
\vspace{-10pt}
The introduced feedback extension is based on a deep learning classifier predicting the user engagement in job interview situations that uses multimodal feature (e.g., gaze, body posture, or gestures) representations of the trainee as input. We exploit the concept of counterfactual explanations to show what the user would need to change to appear more engaged. Therefore, a GAN-driven counterfactual explanation model is trained that transforms those feature representations to corresponding counterfactual explanations, i.e., the feature representations are changed so that the user would have appeared engaged. The explanation generation compares the counterfactual feature vectors with the original feature vectors to derive textual recommendations automatically. Finally, they are presented to the trainee by a socially interactive agent in the role of a job interview coach. \autoref{fig:schema} shows a schematic overview of our approach.

\section{Related work}

Due to the complexity and importance of job interviews, automatic training approaches have been developed to improve the performance of the candidates.
Multiple simulated training systems have been proposed over the years that combine social signal interpretation and virtual agents \cite{baur:2013:jobinterviews,Hoque2013MACHMA,takeuchi:2021:jobinterview}. 
Another example by Gebhard et al. \cite{gebhard:2019:traingjobinterviews} introduced a serious game simulation platform to train social skills. They showed that their training systems can be utilized to teach individuals how to display adequate socio-emotive reactions during job interviews. 
Naim et al. \cite{Naim:2018:automated_analysis_job_interview_performance} introduced a framework for the automatic assessment and analysis of job interview performance. Their proposed system is capable of reliably predicting ratings for interview features such as friendliness, excitement and engagement. Through analysis of the learned feature weights of their regression model they were able to derive general recommendations on how to behave during job interviews, e.g. use filler words less frequently. However, those recommendations are not specific to a situation but rather general guidelines. Takeuchi et al. \cite{takeuchi:2021:jobinterview} developed a job interview training system that provides automatically generated, situation-specific feedback by analysing nonverbal behavior and comparing it to a reference model of ideal nonverbal behavior. The feedback generation was accomplished by defining weights for the shown improper nonverbal behavior in accordance with its importance during the interview. 

Even though providing feedback or guidelines based on weight prioritization may produce satisfactory results, those approaches fail to take the interplay of different shown nonverbal behaviors into account, since each behavior is considered on its own. Imagine a job candidate that is appearing to be low engaged due to a closed body posture with crossed arms and additionally isn't giving his interlocutor much nonverbal feedback like nodding. For such a case it is not enough to consider each behavior or corresponding feature on its own. If we choose to recommend giving more nonverbal feedback, we need to be also aware of how the person is being perceived while changing one of his behaviors. In our case, this would result in a person nodding while still maintaining a closed body posture with crossed arms. Therefore, we argue that it is important to consider the interplay of features when generating personalized feedback and nonverbal behavior recommendations. By utilizing a counterfactual reasoning process we are able to generate feedback that models a holistic recommendation for nonverbal behavior adjustments. This reasoning process tries to answer the question of how should the person have behaved to be perceived as more engaged. For this purpose, the underlying GAN tries to change simultaneously as many features as needed while at the same time trying to change as few features as possible and therefore guaranteeing meaningful recommendations.









\section{Recommendation Generation}
\label{sec:recommendation_generation}

The next sections offer an overview of the different components we implemented to generate behavioral recommendations that point out how the user should have behaved to appear more engaged.

\subsection{Feature Extraction}
\label{feature_extraction}

In order to train a model for engagement recognition and recommendation generation, we modeled a high-level engagement feature set that can be easily interpreted. The feature set consists of 18 metrics mapping facial behavior, body language and conversation dynamics.

During conversations, the face usually occupies most of the interlocutors' attention. A lot of important information regarding the level of engagement can be extracted from the face, respectively the head. In fact, there are multiple studies that found a correlation between head movement / gaze behavior and conversational engagement \cite{Ishii:2010:ESE:2002333.2002339Engagement} \cite{Bednarik:2012:GCE:2401836.2401846Engagement} \cite{10.1007/978-3-642-23974-8_29engagementHeadPose}. Inspired by those findings we defined several features that represent the overall movement of the head and gaze behavior. Moreover, we considered the valence of the face calculated from the facial action units that have been extracted with OpenFace \cite{Baltruaitis2018OpenFace2F}.

Another modality we take into account is the general body language of the job candidate. 
The alignment of the body and the limbs play an important role in broadcasting the state of engagement \cite{bodylanguagecommunication}.
Interlocutors, that are engaged during a conversation, align their bodies to each other, as described in \cite{bodyposture}, ``to create a frame of engagement''.
We tried to cover the general behaviour of the body, as well as specific gestures or poses that are connected to engagement. We defined a group of features that are mainly inspired by the coding system introduced in \cite{bodyposturecodingsystemDael2012}. It contains several metrics to map the orientation and movement of the joints. Those metrics represent amongst other things the overall level of body openness. Besides that, we also calculate a cumulative value over all joints to measure the overall body movement. Lots of body movement may indicate restlessness, which can be an indication for low engagement \cite{d2007posture}.
In addition to that, we also considered the amount of gesticulation an individual performs, as it plays an important role in nonverbal communication \cite{nonverbalcommunication} \cite{bodyposturecodingsystemDael2012}.

Finally, we also covered some form of conversation dynamics. Turn-taking and vocal cues play an important role throughout a conversation \cite{nonverbalcommunicationinhumaninteraction}. During a conversation, the interlocutors usually alternate their speaking turns. Therefore we determine the interlocutor that is currently holding the turn, by considering the general voice activity of the interlocutors. This allows us to draw conclusions about the overall involvement of the individuals during the conversation. An overall low voice activity may imply a conversation with low engaged interlocutors.


\subsection{Engagement Model}
\label{sec:engagement_model}

Based on the feature set introduced in \autoref{feature_extraction} we trained a simple feedforward neural network with two dense layers for the recognition of low and high engagement. For training the network we used the NoXi database \cite{noxicorpus}.
It provides dyadic novice-expert conversations.
We decided on the NoXi corpus since it contains multi-modal multi-person interaction data and its transferability to social coaching scenarios. Moreover, the setup of the corpus allowed for both engaging as well as non-engaging interactions.

 \begin{wrapfigure}{r}{4.6cm}
\includegraphics[width=0.40\textwidth]{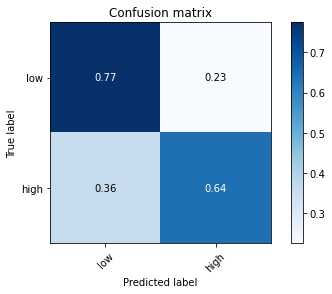}
\caption{Confusion matrix of the neural network for the recognition of low and high conversational engagement (test set).}
\label{fig:confusion_matrix}
\vspace{-30pt}
\end{wrapfigure}
A total of 19 sessions of the NoXi corpus have been annotated regarding conversational engagement resulting in 10.5 hours of training data. The data has been randomly split into training and test sets, so that no sample of the same participant is present in the training and the test set. The training set included 13 sessions and contained 6.8 hours of data. The rest was allocated to the test set. \autoref{fig:confusion_matrix} displays the confusion matrix of the classifier for the test set.


\subsection{Counterfactual Features}
\label{FeatureFactual}

In a next step, to be able to give recommendations on how the user should have behaved to appear more engaged, we apply a counterfactual explanation generation algorithm, i.e., we aim to modify the input feature vectors that were classified as \emph{low engaged} in a way that the classifier would change it's decision to \emph{high engaged}. 
As described in Section \ref{sec:introduction}, the recommendations that we aim for can be seen as counterfactual explanations for the engagement model presented in Section \ref{sec:engagement_model}. 
To generate these counterfactual feature vectors, we used an adversarial learning approach. 
In prior work, Mertes et al. \cite{mertes2021GANterfactual} presented their \emph{GANterfactual} architecture, which extended the CycleGAN framework \cite{zhu2017unpaired}, which is an adversarial approach to domain translation, with further modifications that support the architecture in transforming original samples to counterfactual samples that are classified in a different way by a specific decision system to be explained. To this end, they incorporated the classifier into the training process of their CycleGAN-driven counterfactual explanation system via an additional loss function component. 
For our system, we built a network architecture adapted from the GANterfactual framework, which was originally implemented for generating counterfactual explanations in the image domain. 
The use of the GANterfactual framework has multiple benefits for the recommendation quality: Firstly, the cycle-consistency loss that is an integral part of CycleGANs forces that the learned transformation is minimal, i.e., only relevant features are changed. In the context of recommendation generation, this implies that the generated behavioral recommendations are highly personalized. Secondly, the adversarial loss component that is part of every GAN architecture leads to highly realistic results. Thus, recommendations are not drawn from highly exaggerated or oversimplified feature vectors. Thirdly, the counterfactual loss introduced by Mertes et al. enforces that the counterfactual explanations (in our case, the behavioral recommendations), are valid.
As the engagement model that we used for our system works with feature vectors with no spatial relations between the single features, we replace the convolutional blocks of the original architecture with fully connected blocks. Further, the input layer was adapted to fit the feature representations that we also use for the engagement classifier. The rest of the architecture, as well as the training procedure, was taken from the original GANterfactual framework. For specific technical details, please refer to our implementation.\footnote{Our implementation is available at \url{https://github.com/hcmlab/FeatureFactual}.}
For the GAN-training, we relied on the NOXI dataset, which we also used for training the engagement classifier.
Thus, the adversarial framework learns to convert feature vectors that show low engagement to feature vectors that show high engagement.

\subsection{Textual Recommendations}
After generating the counterfactual feature vectors we compare them to the original feature vectors that represent the shown nonverbal behavior. Depending on the demanded detail of feedback we return the features that had undergone the greatest value transformation. After identifying the most meaningful counterfactual features we convert them into textual feedback. For this purpose, we discretize the features based on a defined textual template. For example, the feature representing the overall activity of the head gets translated into "try to keep your attention on your interlocutor" or "try to use more nonverbal feedback" depending on the present feature value. The amount of discrete classes varies for different features and can easily be adjusted depending on the given use case. The generated feedback is provided verbally to the user by the virtual coach inside the job interview training environment. An example of a recommendation provided by the virtual coach is displayed in \autoref{fig:video}.

\section{Pilot Study}
\label{Study}

The present pilot study's goal was to get preliminary insights about the assessment of a possible job interview training applying GAN driven recommendations. We used a mixed-methods design, combining questionnaires and a semi-structured interview. 
The study was conducted in January 2022. 

\subsection{Method}

\subsubsection{Participants.}
We gathered data from 12 volunteering student participants (7 female, 5 male). 
Participants' age was between 21 and 29 years (\textit{M}\,=\,23.83, \textit{SD}\,=\,2.66). On average, participants attended 4.33 job interviews (\textit{SD}\,=\,2.74; \textit{Min}\,=\,1; \textit{Max}\,=\,10) prior to the study. Two of them had already experience with job interview trainings, three with virtual agents.

\subsubsection{Procedure and Material.}
\begin{figure}[ht]
\centering
\includegraphics[width=.9\textwidth]{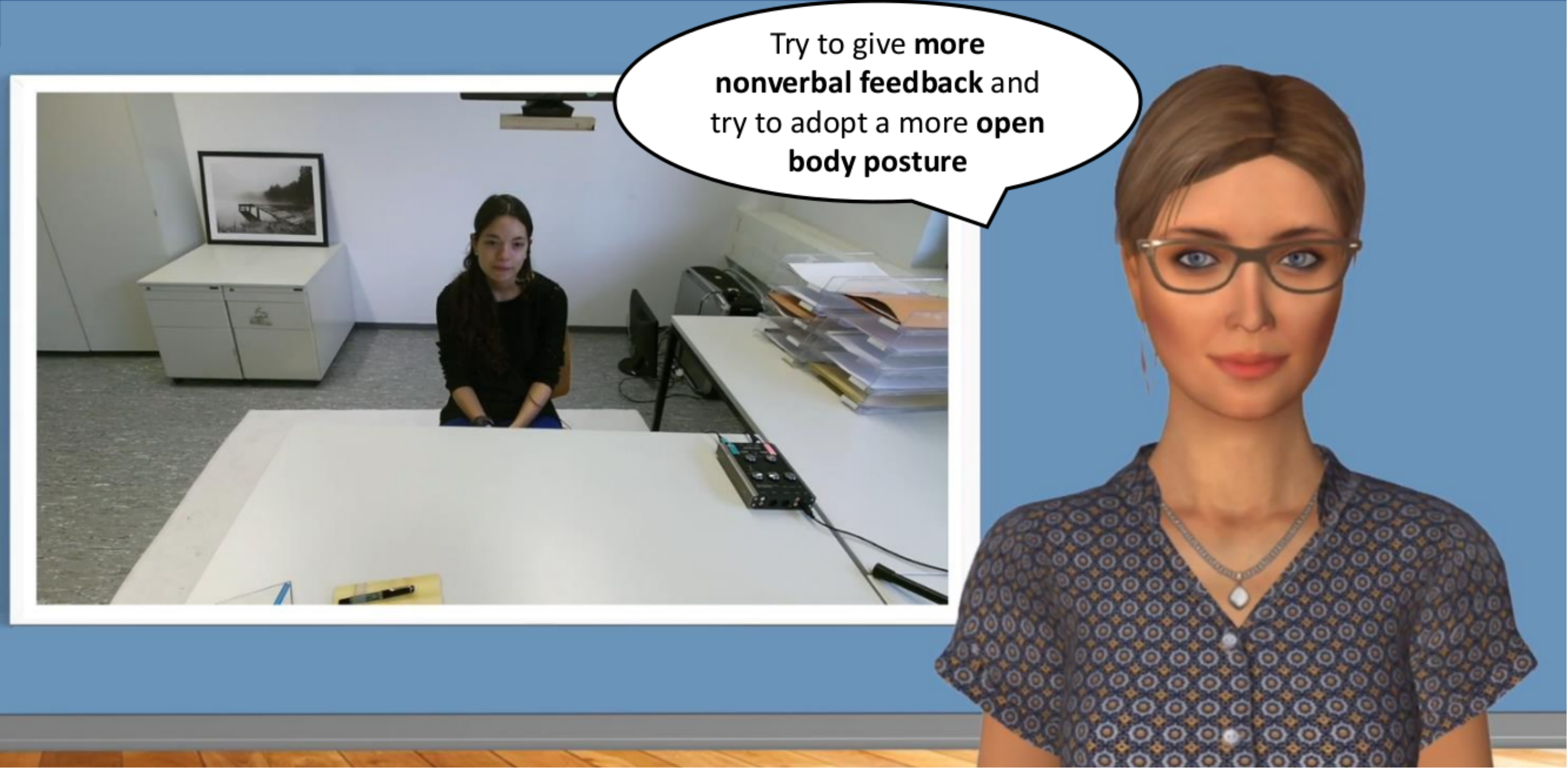}
\caption{Coach giving the recommendation after the mock job interview.}
\label{fig:video}
\end{figure}

In this pilot study, the experimenter and participant met in a video call. After agreeing to the consent form, the experimenter explained the background of the study and presented videos of our job interview training system.
For the videos, we used a multi-modal job interview role-play dataset \cite{schneebergeretal2019} to create behavioral feedback. In that dataset, participants were confronted with a job interview conducted either by an interactive social agent or a human interviewer. Participants were recorded with the MS Kinect2. We used 5 sessions with the human interviewer as input to our job interview training system. The resulting recommendations were then rendered into a video (Fig. \ref{fig:video}) that was shown to the participants. 
The participants saw the part of the job interview training in which the trainee gets the individual feedback from the virtual coach after having a mock job interview. The coach first presents the recorded part of the job interview and gives the recommendation afterwards verbally. Participants were asked to imagine that they were the trainees using the training to practice a job interview.
Next, participants filled in the questionnaires. Then, the semi-structured interview was held. In the end, the experimenter thanked the participants for their participation. The whole procedure took around 25 minutes.

\vspace{-20pt}
\subsubsection{Measurements.}

\textit{Demographics} included age, sex, job interview experience, and job interview training experience.
\textit{Usefulness} was measured with the usefulness scale of the MeCUE \cite{minge2013mecue}. It contains three items. Cronbach's Alpha was .92. 
\textit{Transfer motivation} was measured using four items adapted from \cite{rowold2008evaluation} covering whether training lessons learned will be useful in upcoming situations: ``I believe that my performance in job interviews will improve if I apply the knowledge and skills I have acquired with training.'', ``It is unrealistic to believe that mastering the training content can improve my performance in job interviews. '', ``I can apply skills and knowledge acquired from job interview training to my daily life.'', ``I feel like after the training I could apply the behavior very well. ''. Cronbach's Alpha was .90. 
\textit{Feedback Quality} was measured with four self constructed items: ``I felt the feedback was accurate.'', ``I would have given similar feedback.'', ``I feel like the feedback is helpful.'', ``I don't think the computer can give me accurate feedback.''. Cronbach's Alpha was .87. 
All questionnaire items were answered on a 7 point scale ranging from 1 (\textit{strongly disagree}) to 7 (\textit{strongly agree}).

\noindent The \textit{Semi-structured interview} covered six areas: 1) general impression, 2) persona, 3) other possible use-cases, 4) suggestions for improvement, 5) intention for further use, and 6) added value.

\subsection{Results}


\subsubsection{Questionnaires.}

In the three questionnaires, the following descriptive data was found: 
Usefulness (\textit{M}\,=\,4.72, \textit{SD}\,=\,1.17); Transfer motivation (\textit{M}\,=\,4.92, \textit{SD}\,=\,.94); Feedback (\textit{M}\,=\,4.60, \textit{SD}\,=\,1.26).


\subsubsection{Semi-structured interviews.}

The answers gathered in the semi-structured interview were analyzed and categorized for each of the six areas separately:

\noindent 1) Regarding the \textit{General impression}, participants mentioned six times that the recommendations were useful / feasible (e.g., ``Simple tips that were easy to implement, but have a big impact.'') or comprehensible (2x). Three participants mentioned that the recommendations were too unspecific. Once each was mentioned that the recommendations are not useful (``Would prefer feedback on the content of my answer. Job interview is too stressful for me such that I could focus on non-verbal behavior.'') and too obvious (``If I saw myself in the video, I would have known that I have to improve the recommended behaviors.''). 



\noindent 2) Participants described the \textit{persona} as someone with a wish to improve (7 namings) that is open for new thing (3 namings), career oriented (2 namings), young (2 namings), self reflective (2 namings) or non-self reflective (1 naming).

\noindent 3) As \textit{other possible use-cases} participants named training to improve communication skills	in general (8 namings) and for more specific groups, like patients with anxiety disorders or people with social phobias. The named also other possible situations like preparing for challenging employee appraisals, conflict resolution dialogs, or other high stakes situations. Another named use-case was public speaking (4 namings). 

\noindent 4) Participants mentioned seven times that they would like to have more specific recommendations, e.g. “The agent could say something like: Nonverbal feedback is nodding, for example." Moreover, they thought that recommendations based on the content of the answers would be helpful (2 namings). Also, some participants noted that the agent could be improved (3 namings), like using a more empathic voice. One participant noted that an interactive training mode, where you practice recommendations directly and get instant feedback would be helpful.


\noindent 5) \textit{Intention for further use} was indicated by 9 participants. Three could not imagine using the training.

\noindent 6) The \textit{added value} of the training was for most of the participants that the recommendations are given directly on a specific behavior shown in a specific situation during the job interview. Moreover, one participant mentioned that the training was especially helpful as it gives a low-threshold possibility to practice job interviews that could be offered by agencies supporting people to find employment. One other participant said that having an agent instead of a human giving recommendations decreases the feeling of being judged for mistakes.





\subsubsection{Recommendation generation}
\begin{figure}[ht]
\centering
\includegraphics[width=1\textwidth]{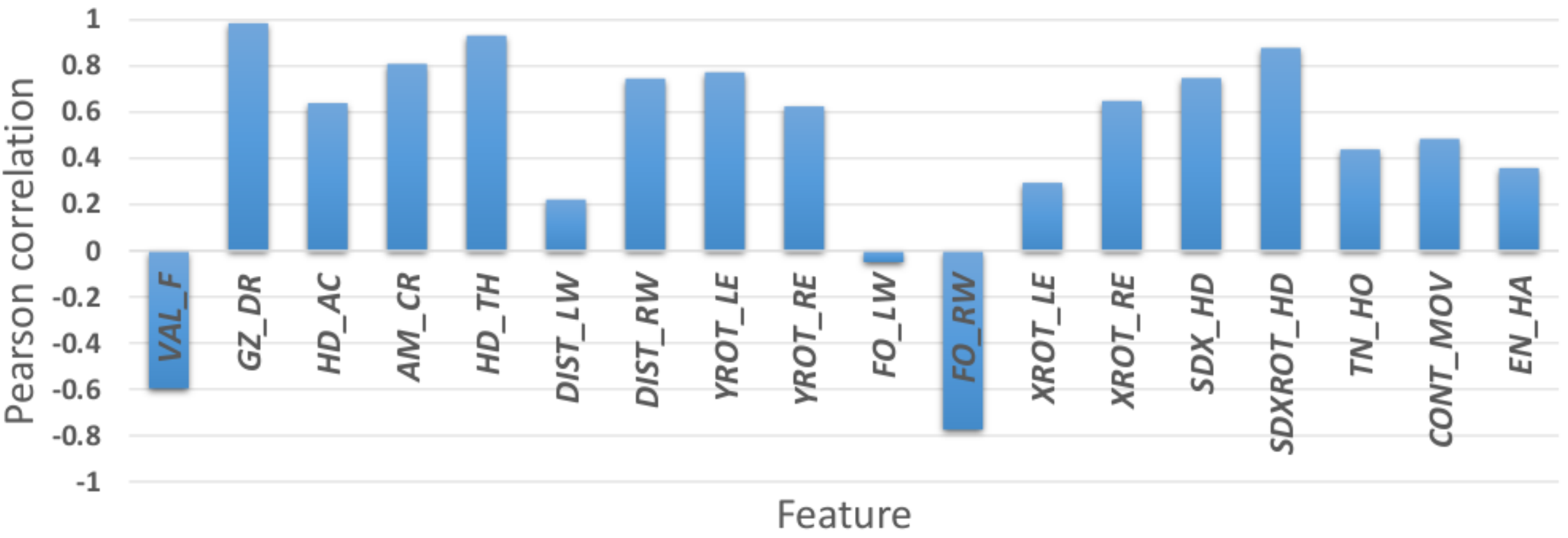}
\caption{Pearson correlation between the absolute change of the feature values and the LIME classification relevance scores for every feature. The features are from left to right: \textit{Valence Face, Gaze behavior, Head activity, Arms crossed, Head touch, X distance of left/right wrist and hip, Y rotation left/right elbow, Y distance of left/right wrist and hip, X rotation left/right elbow, Standard deviation head movement in X axis, Standard deviation Head X rotation, Turn hold, Continuous movement, Gesticulation}.}
\label{fig:pearson_correlation}
\end{figure}

As described in \autoref{FeatureFactual} we incorporated a classifier for the recognition of engagement into the training process of the GAN via an additional loss function component. In order to verify the validity of our approach, we examined whether the counterfactuals generated by the GAN are modifying the features that the engagement classifier identified as important for the classification of low and high engagement. 
For this evaluation, we used five sessions of the multi-modal job interview role-play dataset \cite{schneebergeretal2019} that have also been used in \autoref{Study} and extracted the importance scores of every feature in regard to the model's classification with LIME \cite{lime}.
Next, we calculated the absolute value change of how much each feature has been modified by the counterfactual transformation. Afterwards, we calculated the Pearson Correlation Coefficient between the importance scores of every feature and the absolute change of each feature, see \autoref{fig:pearson_correlation}. High correlation scores indicate that the counterfactual feature transformation is in line with the corresponding importance of the feature. The more important a feature is for the classification of a sample the greater also should be the change of the feature in order to result in a different classification result. Seven features showed a strong positive correlation (GZ\_DR, AM\_CR, HD\_TH, DIST\_RW, YROT\_LE, SDX\_HD, SDXROT\_HD), six features had a moderate positive correlation (HD\_AC, YROT\_RE, XROT\_RE, TN\_HD, CONT\_MOV, EN\_HA) and two features presented with a low positive correlation (DIST\_LW, XROT\_LE). Moreover, FO\_RW had a strong negative correlation, VAL\_F showed a moderate negative correlation and FO\_LW had a weak negative correlation. 

Moreover, we conducted a computational evaluation to investigate how well the generated counterfactual features change the decision of the engagement classifier. For this evaluation, we also used the multi-modal job interview role-play dataset. We found that 96.49\% of the generated counterfactual feature vectors led to a different decision of the engagement model as the original input features.



\section{Discussion and Conclusion}
We introduce a novel approach for generating textual nonverbal behavior recommendations in job interview training environments. In a pilot study, we presented the approach to participants. The results indicate that such training could be helpful to prepare for job interviews successfully. The recommendations given by the system were found to be helpful and comprehensible, and transferable to other use cases. Moreover, most participants noted that the proposed approach adds additional value to the training by giving recommendations directly on a specific behavior in a specific situation. Part of the underlying training system automatically extracts situations that could be improved and displays them alongside the recommendation presented by the virtual coach. However, the pilot study also revealed that the recommendations should be more specific. Therefore, in future work, the template used for discretizing the counterfactuals should be extended to be more diverse and specific or use natural language processing to generate textual recommendations from counterfactuals directly. The latter would need additional annotation and training effort.


Moreover, we examined the validity of our GAN-driven recommendation generation approach by calculating the Pearson correlation coefficient between the absolute changes of the feature values after counterfactual transformation and the importance of the features the classifier attributed to them regarding the classification result. We showed that most of the features (15 out of 18 features) had a moderate to strong correlation, which emphasizes the validity of the proposed approach. Only the two features corresponding to the relative position and movement of the left wrist and the feature representing the flexion of the left elbow presented a weak correlation. 

Further, it is interesting to point out that the feature representing the relative movement of the right wrist (FO\_RW) has shown a strong negative correlation. This means that the counterfactual suggests decreasing the relative distance from the wrist to the rest of the body when the current feature value is an indication for low engagement. The opposite is the case when the current feature value indicates being highly engaged, here the relative distance should be increased. This indicates that for the given job interview data, the engagement classifier attributes a lower wrist distance towards the body as appearing higher engaged. A similar case presented itself for the valence of the face. For this feature, we found a moderate negative correlation. For the valence of the face, the classifier interprets lower valence values, meaning a more serious facial expression, as a sign for higher engagement. This interpretation is most likely related to the dataset used for training the classifier and the corresponding conversational engagement annotations. Therefore, extending the used training data for both the classifier and the GAN for future work makes sense. Especially the classifier might benefit from more training data as the accuracy scores leave room for improvement. Also, the current classifier only distinguishes between low and high engagement. It would also be interesting to investigate the resulting counterfactuals when using a more fine-grained representation for conversational engagement. Further, we also investigated how well the generated counterfactual features can change the decision of the engagement classifier. Overall, 96.49\% of the counterfactual feature vectors led to a different decision of the engagement classifier as the original input features.
This indicates that our GAN-driven approach enables to generate recommendations that, when being adopted, are consistently leading to a perception of high engagement. 
The computational evaluation, as well as the user study, indicate that the generated recommendations are valid and helpful in the context of job interview coaching scenarios.
\subsubsection{Acknowledgments.} 
This work presents and discusses results in the context of the
research project ForDigitHealth. The project is part of the Bavarian Research Association on Healthy Use of Digital Technologies and Media
(ForDigitHealth), funded by the Bavarian Ministry of Science and Arts. Further, the work described in this paper has been partially supported by the BMBF under 16SV8688 within the MITHOS project, and by the BMBF under 16SV8493 within the AVASAG project.
\bibliography{main}
\bibliographystyle{splncs04}

\end{document}